# Quantum Coherent Transport in SnTe Topological Crystalline Insulator Thin Films

B.A. Assaf[1], F. Katmis[2,3], P. Wei[3], B. Satpati[4], Z. Zhang[5], S.P. Bennett[6], V.G. Harris[6], J.S. Moodera[2,3] and D. Heiman[1]

[1] Dept. of Physics, Northeastern University, Boston, MA 02115, USA
[2] Francis Bitter Magnet Lab, MIT, Cambridge, MA 02139, USA
[3] Dept. of Physics, MIT, Cambridge, MA 02139, USA
[4] Saha Institute of Nuclear Physics, 1/AF Bidhannagar, Kolkata 64, India
[5] Advanced Photon Source, Argonne National Laboratory, Argonne, IL 60439, USA
[6] Dept. of Electrical and Computer Engineering, Northeastern University, Boston, MA 02115, USA

**Abstract:** Topological crystalline insulators (TCI) are unique systems where a band inversion that is protected by crystalline mirror symmetry leads to a multiplicity of topological surface states. Binary SnTe is an attractive lead-free TCI compound; the present work on high-quality thin films provides a route for increasing the mobility and reducing the carrier density of SnTe without chemical doping. Results of quantum coherent magnetotransport measurements reveal a multiplicity of Dirac surface states that are unique to TCI. Modeling of the weak antilocalization shows variations in the extracted number of carrier valleys that reflect the role of coherent intervalley scattering in coupling different Dirac states on the degenerate TCI surface.

Topological crystalline insulators[1,2] (TCI) are a class of topological insulators[3,4,5,6,7,8], whereby a band crossing protected by crystalline mirror symmetry gives rise to nontrivial topological surface states. In (Pb,Sn) chalcogenides,[9,10,11,12,13,14] the TCI state exists as a result of the mirror symmetric character of the rocksalt structure of these compounds. As a consequence, there exists 4 Dirac cones at the surface of SnTe. These 4 cones merge into two Dirac cones [15,16] as the Fermi level moves deep into the valence band and bulk states become populated. This makes SnTe particularly unique as a platform to study valley-degenerate topological systems in magnetotransport experiments. However, the growth of high-quality SnTe thin films is a prerequisite to study such degeneracy effects, and is also vital for future devices based on TCI.

Binary SnTe is known to exhibit heavy *p*-type conductivity as a result of Sn vacancies and Te substituting for Sn.[17,18,19] A second hurdle is the fact that at small thicknesses (below 1 μm) films tend to be highly granular and rough[20], which can significantly reduce the carrier mobility. Most of the recent work on SnTe and $Pb_{1-x}Sn_xTe$ has been focused on micron-thick films, or the use of buffer layers in order to avoid the problems associated with these issues. [21,22,23,24,25]

In this work, nanometer thick films of SnTe were grown by molecular beam epitaxy (MBE) on insulating $BaF_2$ (001) substrates. We find that increasing the growth temperature from 220 to 340 °C, not only improves the crystalline quality and morphology of the films, but also leads to a decrease in the carrier concentration and hence reduces the Fermi level. Hall mobilities up to $\mu$=760 cm$^2$/Vs were obtained with a corresponding carrier density of $p$=8×10$^{19}$ holes/cm$^3$. These improvements lead to the observation of a robust cusp in the low-field magnetoconductivity, known as weak antilocalization (WAL). Analysis of the WAL indicates the likely presence of multiple coherently-coupled carrier valleys, which are manifestations of the topological surface states. Figure 1(a) illustrates the 4 Dirac cones of the SnTe (100) surface states along the $\overline{\Gamma} - \overline{X}$ lines of the first Brillouin zone.

The SnTe films were grown by MBE on (001)-oriented insulating $BaF_2$ substrates with thicknesses of 30 to 60nm. The substrates were preheated at 340°C for 2 hours prior to the growth.

High-purity elemental Te and Sn were evaporated from Knudsen cell and electron gun sources, respectively. The substrate temperature was varied between 220°C and 340°C and the pressure during the growth was less than $5\times10^{-9}$ torr. The growth rate for films considered in this work was fixed at 1.2 nm/min. The Sn to Te ratio was kept on the Te rich side to avoid the formation of Sn clusters. Films were capped in-situ with 4nm of $Al_2O_3$. Cross-sectional transmission electron microscopy (XTEM), synchrotron X-ray diffraction (XRD), atomic force microscopy (AFM), and magnetotransport measurements confirmed the high film quality when grown above 300°C.

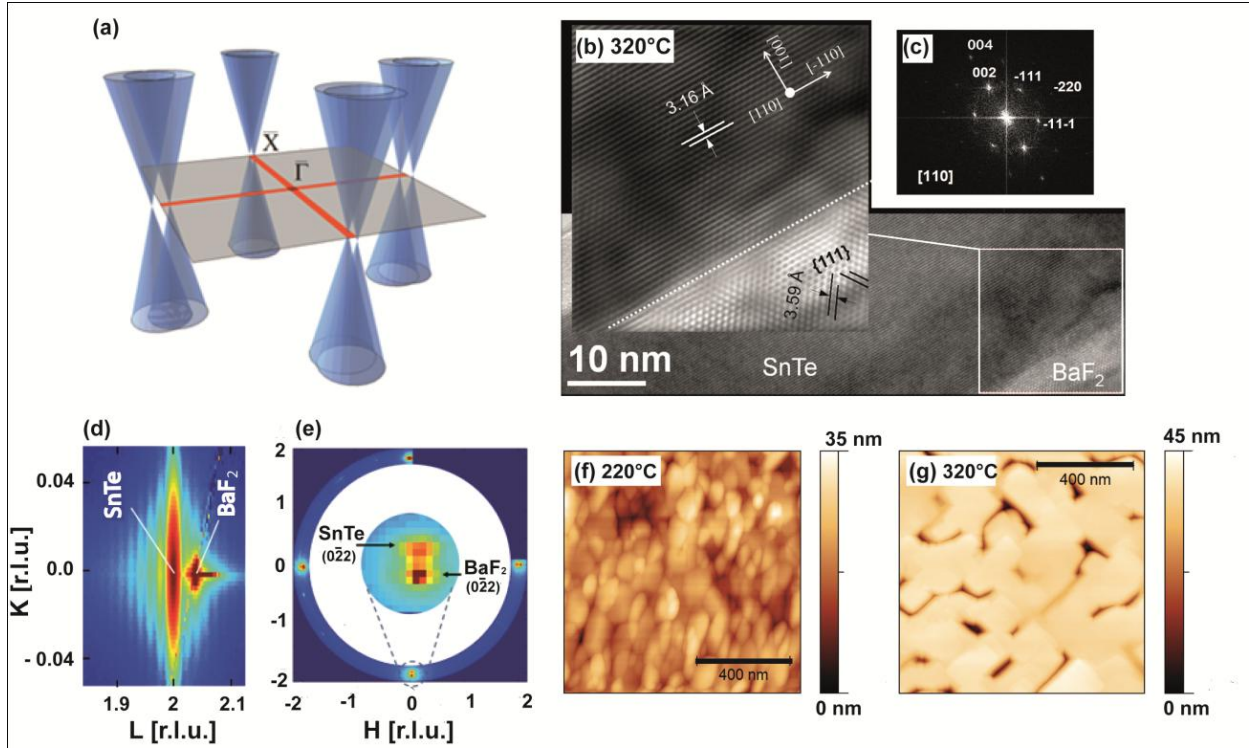

FIG. 1: (Color online) (a) SnTe (001) surface states illustrated as 4 Dirac cones along the $\bar{\Gamma}-\bar{X}$ lines of the first Brillouin zone. (b) XTEM images of a SnTe film grown on $BaF_2$ (001) at 320°C, and a Fourier-filtered image of the interface between the film and substrate. (c) Fourier pattern of the interface. (d) Reciprocal space-map of the (002) XRD Bragg peak of SnTe. (e) $\phi$-scan showing the 4-fold {022} XRD Bragg series for SnTe and $BaF_2$, with an expanded view of one peak in the center. (f-g) AFM images of films grown at 220 and 320°C, respectively.

Figure 1(b) shows the XTEM image of a high-quality film grown at 320°C. Clear atomic planes can be resolved in the bulk, all the way down to the substrate interface, as seen in the Fourier-filtered image on the left. The interface appears very smooth and no misfit or threading dislocations are observed in the TEM image. The FFT pattern in Fig. 1(c) shows no detectable spot separation, which further confirms the formation of a lattice-matched interface. Figure 1(d) shows a reciprocal space map of the *K-L* cut near the (002) XRD Bragg peak for a film grown at 320°C. A strong SnTe peak is seen alongside the substrate peak. The *c*-axis lattice constant was found to be $c = 6.324\pm0.003$Å from XRD in agreement with the value measured in the TEM image, $c = 6.36\pm0.06$Å. A $\phi$-scan was also performed for the {022} Bragg reflections, shown in Fig. 1(e), and reveals the existence of a four-fold symmetric crystal structure in the film that aligns with the $BaF_2$ substrate crystal structure. Overall, the XRD study confirms the formation of a high-quality, epitaxial, 40nm thick SnTe film on $BaF_2$ (001). The film morphology was also greatly improved at higher growth temperatures, as seen in the AFM images in Figs. 1(f-g). In the film grown at 220°C, the surface was highly textured and exhibited island-like features that are no larger than 50nm. For growth at 320°C wider and larger plateaus can be seen extending more than 500 nm in width and length.

The Hall effect and the magnetoresistance were measured at 2K and up to 5T using a modified probe[26] inserted in a Quantum Design MPMS magnetometer. The Hall conductivity, $\sigma_{xy}$, was extracted by inverting the Drude resistivity tensor[27,28]. As seen in Fig. 2(a), the slope of $\sigma_{xy}$ increased as the growth temperature was increased between 220 and 290°C, and at 320°C further increased and became highly non-linear. The non-linearity was corroborated up to $B$=9T. Although, it has been argued that the coexistence of surface-states and bulk-states contributing to the transport in Bi-based TIs may give rise to a non-linear Hall conductivity[29,30], the bulk itself may exhibit a nonlinearity when the mobility of bulk electrons approaches the limit $(\mu B)^2 \sim 1$.[31] In that limit, the full Drude $\sigma_{xy}$ is given by

$$\sigma_{xy} = \frac{e n_H \mu_H^2 B}{1 + \mu_H^2 B^2}. \tag{1}$$

Here, $n_H$ is the Hall carrier density and $\mu_H$ is the Hall mobility. In the limit where $(\mu_H B)^2 \ll 1$, one has the conventional linear Hall conductivity given by $\sigma_{xy} = e n_H \mu_H^2 B$. However, as $(\mu_H B)^2 \to 1$, the full form of the Drude Hall conductivity must be used, which then results in a non-linear Hall conductivity. The results of this fitting are shown by the solid curves in Fig. 2(a). The Hall carrier density and mobility were extracted for 5 films grown at different temperatures. In addition to these 40nm thick films, 3 films with thickness of 30, 40 and 60nm were grown at 320°C and measured.

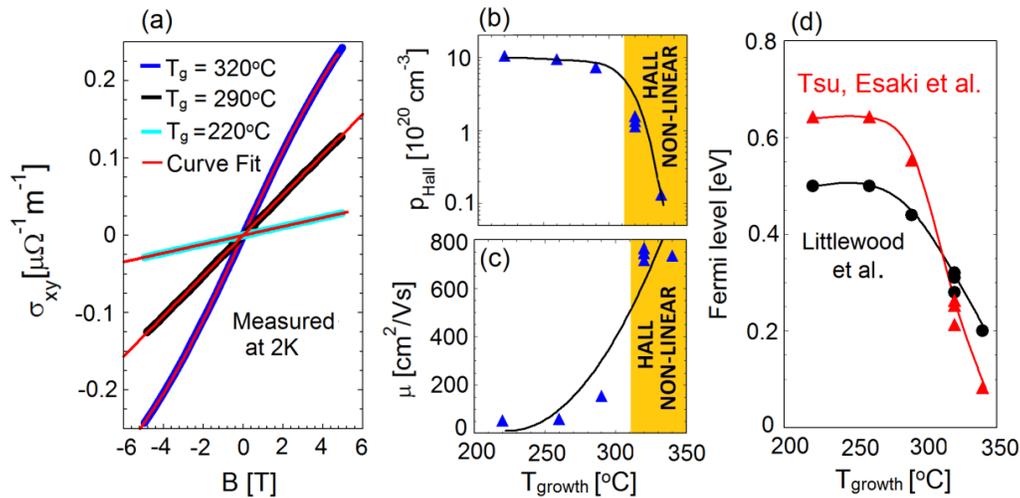

FIG. 2: (Color online) (a) Hall conductivity versus magnetic field for 40nm thick films grown at 220, 290 and 320°C. Solid red curves are fits to the Drude Hall conductivity. Growth-temperature dependence of: (b) Hall carrier density showing 2 orders of magnitude decrease; and (c) Hall mobility showing one order of magnitude increase. The solid curves are guides for the eye. (d) Fermi energy extracted from the Hall carrier density for films grown at different temperatures, using the results of Tsu et al.(red)[19] and Littlewood et al. (black)[17].

The carrier concentration and mobility were found to be strong functions of growth temperature. Figures 2(b), (c) and (d) show that increasing the growth temperature from 220 to 340°C resulted in: (*i*) the Hall carrier density decreasing by 2 orders of magnitude from $1.0 \times 10^{21}$ to $8 \times 10^{18} cm^{-3}$; (*ii*) the Hall mobility increasing by an order of magnitude from 60 to 720 $cm^2$/Vs; and *(iii) the Fermi*

*level,[32] measured from the center of the band gap, decreasing by more than a factor of two*. As SnTe is always *p*-type due to Sn vacancies and Te-antisites, an increased substrate temperature may destabilize the formation of Te-antisites and thus reduce the number of acceptor states that leads to a decreasing Fermi level. Interestingly, this behavior is in qualitative agreement with what was recently predicted in SnTe,[33] when the ratio of Sn to Te was varied. Similar trends in the carrier density were also observed in highly *p*-type GeTe thin films.[19]

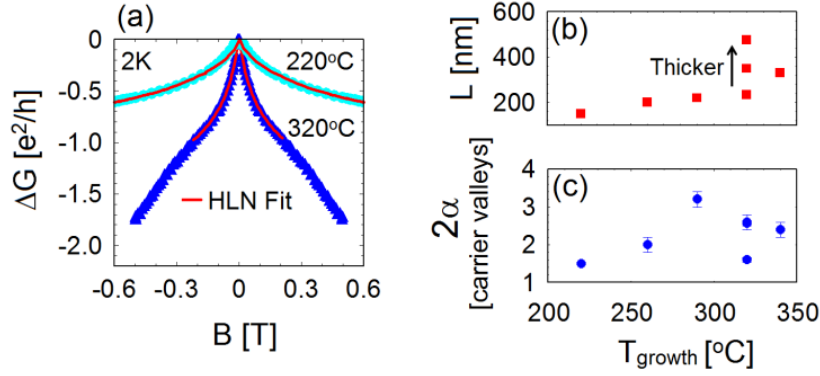

FIG. 3: (Color online) Results of WAL conductance of SnTe films at 2K. (a) Change in conductance versus magnetic field showing WAL cusps in films grown at 220°C (light blue) and 320°C (dark blue). Blue marks are experimental data, and solid red curves are fits to the HLN model. Growth-temperature dependence of: (b) phase coherence length, *L*; and (c) number of carrier valleys, $2\alpha$. The error bars in (c) represent the uncertainty in fitting the WAL for a factor of 3 range in fields, from 0.1 to 0.3 T.

Next, we study quantum coherent WAL in SnTe films, as extracted from the low-field magnetoresistance (MR) at 2K. WAL arises as a result of spin-momentum locking at the surface of SnTe, and we will later argue that it is unlikely for the bulk bands of SnTe to give rise to such an effect. Fitting our experimental data with the Hikami-Larkin-Nagaoka (HLN) model[34], typically used for topological surface states[35,36,37,38,39,40], allows us to extract the coherence length *L*, as well as $2\alpha$ the number of coherent transport channels. In valley degenerate semiconductors,[41] $2\alpha$ is also equal to the number independent carrier valleys $2\alpha$ ($\alpha=1/2$ per valley) contributing to the transport.[42] The change in sheet conductance versus magnetic field B in the HLN model is given by[42,43]

$$\Delta G = -\frac{\alpha e^2}{\pi h}\left[\psi\left(\frac{\hbar}{4eL^2 B}+\frac{1}{2}\right)-\ln\left(\frac{\hbar}{4eL^2 B}\right)\right]. \quad (2)$$

Figure 3(a) shows the 2K conductivity data and model fitting for films grown at 220 and 320°C. The differences illustrate a general feature, where the WAL cusp is reproducibly narrower and stronger for films with higher mobility (grown at higher temperature). Figures 3(b,c) show values for *L* and $2\alpha$ at 2K extracted for all films, including those grown at temperatures between 220 and 320°C, as well as 3 different thicknesses grown at 320°C. A systematic increase in *L* is observed with increasing growth temperature, reflecting the increase in mobility that is related to the improved film morphology. The coherence length thus increases with increasing grain size. The trend in $2\alpha$ is however more complex and is the subject of the remainder of the discussion. Although an initial increase in $2\alpha$ is observed up to a growth temperature of 290°C, no clear trend is observed beyond this temperature when plotted in this way. As $2\alpha$ allows us to indirectly probe the Fermi surface, we consider the behavior of $2\alpha$ as a function of the Fermi level $E_F$ extracted using ref. 17. In this way we gain insight into the electrical contributions from different bands. Earlier, we extracted the Fermi level - measured from the center of the gap - from the carrier density. Decreasing the Fermi level by varying the growth conditions, without chemical doping, is highly advantageous for investigating the topological properties of SnTe.

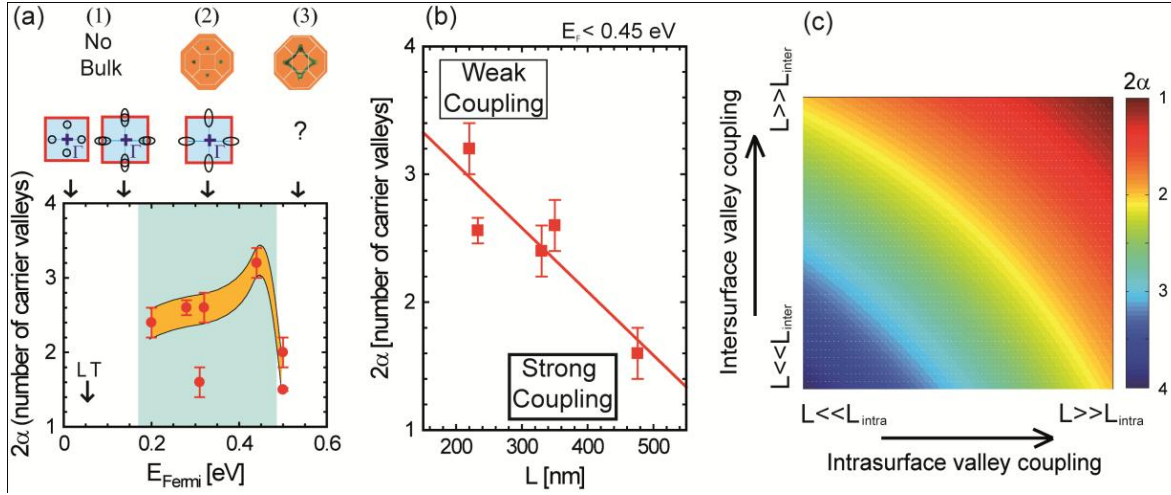

FIG. 4: (Color Online) Changes in the Fermi-surface states in SnTe films from analysis of WAL conductance at 2K. (a) Plot of 2α (number of carrier valleys) as a function of Fermi level. Data points are shown in red. Trend in orange is a guide for the eye. The Lifshitz transition (LT) occurs at $E_F$ = 0.05 eV. Diagrams above the plot illustrate 3 different WAL regimes at different Fermi levels corresponding to 3 different Fermi surfaces for surface states (blue squares) and bulk states[17] (orange octagons). (b) Number of carrier valleys versus $L$ for Fermi energies below 0.45 eV, showing a linear relation between number of carrier valleys and phase coherence length. (c) Qualitative variation of 2α in regime (2) as a function of the strength of valley coupling for both intrasurface and intersurface valley coupling.

Figure 4(a) plots the number of carrier valleys (2α) extracted from the HLN model as a function of Fermi level. We first point out that many of the data points lie higher than 2α=2. A value of 2α greater than 2 is indicative of more than one Dirac surface valley per surface contributing to the transport, as 2α=2 is the maximum number of states for one Dirac surface valley and two surfaces. (The single outlier at $E_F$=0.3eV is considered later.) At larger Fermi level, $E_F$>0.5eV, 2α decreased to below 2α=2. We now distinguish three different transport regimes as a function of Fermi level, labeled (1), (2) and (3). In regime (1) at low $E_F$, the Fermi surface consists of four Dirac surface valleys and no bulk states; a maximum of 2α=8 is thus expected for two surfaces. As the Fermi level moves deeper into the valence band, a Lifshitz transition at $E_F$=0.05eV causes pairs of Dirac surface valleys to merge into four pairs of concentric constant energy contours.[10,15,16] Below the Lifshitz transition in the valence band, an inner Dirac cone and an outer Dirac cone having opposite chirality coexist.[13] The quantum coherent transport in this case is not yet understood. As bulk states become more and more populated, the inner Dirac cone is "swallowed" by trivial bulk states, thus reducing the total number of Dirac valleys by a factor of two. Regime (2) emerges between $E_F$≈ 0.2 and 0.5eV,[13] having a maximum of 2α=4 Dirac surface valleys, and a likely coexistence of surface and bulk states. Finally, in regime (3) when the Fermi level exceeds 0.5eV in the valence band, the shape of the bulk Fermi surface undergoes a transition whereby the 4 bulk *L*-valleys merge into a large tubular Fermi surface, shown in the upper right illustration.[17,18] The behavior of the topological surface states at such deep Fermi levels is not known, but the drop in 2α at 0.45eV may be attributed to a similar change in the shape of the surface valleys. Our films dominantly belong to regime (2). Most noteworthy though, is that the values for 2α fall between 2.2 and 3.4 carrier valleys in this regime, where we should be able to resolve a maximum of 4 surface valleys. The experimental spread in the measured number of carrier valleys in this regime, below 0.45eV, is a result of coherent valleys coupling.

Figure 4(b) plots the number of carrier valleys (2α) as a function of phase coherence length. Here we see a clear trend where 2α decreases for increasing *L*. This trend includes the apparent outlier in Fig. 4(a) at $E_F$=0.3eV that has the longest *L* and a small 2α. The linearly-decreasing behavior can be explained by considering weak and strong coupling between the valleys. In SnTe, two types of coherent valley coupling can occur, *inter* and *intra*surface scattering. As in the case of surface-to-bulk coupling in $Bi_2Se_3$,[35,42] *inter*surface valley coupling can occur in SnTe when a charge carrier is able to scatter between the top and bottom surfaces via the bulk while remaining coherent. The bulk thus plays an essential role in coupling the top and bottom surface. In addition, given the valley degeneracy of SnTe,[2] *intra*surface valley coupling can occur, when a carrier is able to scatter between two Dirac valleys located on the same surface while remaining coherent. This is similar to what is typically observed in graphene[44,45] and Si inversion layers.[41,46] Accordingly, a longer coherence length should result in stronger valley coupling, and thus a smaller experimental 2α. Our data strongly reflects this fact, as 2α increases with decreasing *L* in regime (2). We are thus able to resolve at least 3 independent carrier valleys for the shortest coherence length where coupling is weakest. This is potential transport evidence of the degenerate TCI surface band structure, yielding 3<2α<4 in regime (2). In view of these results, we find that valley coupling – resulting from both intersurface coupling via the bulk and intrasurface coupling due to Dirac band degeneracy – is able to account for the complete picture for films behaving according to regime (2).

It is instructive to draw a qualitative picture of how 2α varies in regime (2), below the Lifshitz transition, as a result of the coexistence of inter and intrasurface scattering mechanisms. This is summarized in Fig. 4(c). Here, the intrasurface and intersurface valley scattering lengths are referred to as $L_{intra}$ and $L_{inter}$, respectively. A total of 4 Dirac carrier valleys can exist in regime (2),[10,2] which may be all resolved in the absence of valley coupling, leading to 2α=4 when $L<<\{L_{inter}, L_{intra}\}$. For weak intersurface valley coupling, increasing the intrasurface valley coupling will lead to 2α continuously decreasing from 4 to 2.[41,42,46] In the fully intrasurface-coupled regime, where ($L<<L_{inter}$ and $L>>L_{intra}$), 2α=2. An equivalent scenario arises when exchanging inter and intrasurface couplings. In the fully intersurface-coupled regime, where ($L>>L_{inter}$ and $L<<L_{intra}$), 2α=2.[35,42] When both intersurface and intrasurface coupling are strong, where $L>>\{L_{inter}, L_{intra}\}$, 2α=1. Figure 4(c) show the qualitative behavior of 2α as $L_{inter}$ and $L_{intra}$ are varied independently. It is interesting that, in contrast to standard TIs, a TCI is expected to exhibit multivalley behavior even in a partially-coupled regime. Values of 2α greater than 2, even approaching 4 are thus expected. This reality is strongly reflected in our data where the number of valleys is consistently greater than 2, and even larger than 3.

Finally, it is worthwhile to examine possible contributions to the WAL that might arise from bulk bands, all the while keeping in mind that WAL is a dominant 2D effect. WAL can arise in a trivial semiconductor if Rashba-split 2D bands exist at the surface.[47,48,49,50] Such spin-orbit split bands require a surface inversion layer that can develop from impurity adsorption at the surface.[51] An inversion layer is not likely here for several reasons: our films are capped with 4 nm of $Al_2O_3$, the SnTe (001) surface is neutral, and the high carrier density and large dielectric constant rapidly screen out unwanted surface charging. The rapid screening in SnTe also limits band bending to a few meV.[27,51,52] The bulk bands of

SnTe are thus unlikely to yield WAL. WAL is thus viewed as a consequence of spin-momentum locking of the TCI surface states, but the bulk bands play an essential role in coupling the top and bottom surfaces.

In conclusion, high-quality thin films of SnTe were grown on $BaF_2$(001) for electrical transport studies. Higher growth temperatures greatly improved the crystal quality and film morphology, which led to an increase in Hall mobility and a large decrease in carrier concentration. Thus, we were able to tune the Fermi level between 0.2 and 0.5eV below the gap, an important realization for future work on the TCI properties of SnTe. This variation in Fermi level facilitated magnetoconductance WAL experiments, revealing a trend related to a changing Fermi surface topology at deep Fermi levels. Our observation of up to 4 carrier valleys is a likely consequence of the unique multiplicity of Dirac cones in a TCI, unlike the single Dirac cone in standard Bi-based TI. It was also found that the observed number of independent carrier valleys decreased as the phase coherence length increased – a result of increased intervalley coupling. The coexistence of intersurface and intrasurface valley coupling results in complex variations in the number of carrier valleys measured in WAL. The present results encourage theoretical work and future experiments on intervalley coupling and quantum coherent transport in TCI materials.

### Acknowledgements

We thank L. Fu, H. Steinberg, V. Fatemi, I. Zeljikovic and C. M. Schlepütz for helpful discussions and T. Hussey for technical assistance. We also thank M. Jamer and T. Devakul for comments on the manuscript. The work is supported by DMR-0907007 from the National Science Foundation. J.S.M., F.K. and P.W. also acknowledge support from NSF-DMR-1207469, ONR-N00014-13-1-0301, and MIT MRSEC through NSF under award No. DMR-0819762. Part of this work was carried out at the Advanced Photon Source that is supported by the U. S. Department of Energy, Office of Science, Office of Basic Energy Sciences, under Contract No. DE-AC02-06CH11357.